\documentclass[prb,aps,twocolumn,eqsecnum,superscriptaddress,showpacs,amssymb]{revtex4}
\usepackage{graphicx}
\usepackage{dcolumn}
\usepackage{bm}
\usepackage{amsmath,amssymb}
\usepackage{psfrag,subfigure}
\def\beq{\begin{equation}}
\def\eeq{\end{equation}}
\def\bea{\begin{eqnarray}}
\def\eea{\end{eqnarray}}
\def\nn{\nonumber}

\begin{document}

\title{Ice: a strongly correlated proton system}

\author{A. H. Castro Neto}
\affiliation{Department of Physics, Boston University, 590 Commonwealth Ave., Boston, MA 02215, USA}
\author{P. Pujol}
\affiliation{Laboratoire de Physique, Groupe de Physique 
Th{\'e}orique de l'{\'E}cole Normale Sup{\'e}rieure, Lyon, France}
\author{Eduardo Fradkin}
\affiliation{ Department of Physics, University of Illinois at Urbana-Champaign, Urbana, IL 61801-3080, USA. }
\date{\today}

\begin{abstract}
We discuss the problem of proton motion in Hydrogen bond materials
with special focus on ice. We show that phenomenological
models proposed in the past for the study of ice can be recast
in terms of microscopic models in close relationship to the ones
used to study the physics of Mott-Hubbard insulators. 
We discuss the physics
of the paramagnetic phase of ice at $1/4$ filling (neutral ice)
and its mapping to a transverse field Ising model and also to
a gauge theory in two and three dimensions. We show that
H$_3^+$O and HO$^-$ ions can be either in a confined or
deconfined phase. We obtain the phase diagram of the problem
as a function of temperature $T$ and proton hopping energy $t$
and find that there are two phases: an ordered insulating phase
which results from an order-by-disorder mechanism induced by
quantum fluctuations, and a disordered incoherent metallic phase 
(or plasma). We also discuss the problem of decoherence in the 
proton motion introduced by the lattice vibrations (phonons) 
and its effect on the phase diagram. Finally, we suggest
that the transition from ice Ih to ice XI observed experimentally
in doped ice is the confining-deconfining transition of our
phase diagram.
\end{abstract}


\maketitle

\section{Introduction}
\label{introduction}

Hydrogen bonds (or H-bond) are ubiquitous in physics, chemistry, and biology.
The nature of H-bonds was studied in great detail by Pauling
who predicted the mixed chemistry of H-bonds \cite{pauling}, namely, 
H-bond shares characteristics of ionic and covalent bonds.
On the one hand, a water molecule has a large dipole moment due
to the electronegative character of the O atom and therefore
water molecules are attracted to each other.
This is the classical aspect of the H-bonding. On the other hand,
the sigma bonding between H and O is strongly covalent with
clear quantum mechanical nature. In the process of formation of
an ice crystal electrons from the sigma bond can be shared by
two water molecules leading to a strong link between them.
Compton scattering experiments have confirmed the quantum
nature of H-bonds in ice crystals \cite{cova}.
It has been clear since the early experiments in H-bond systems
that although the physics of electrons in ice is important
\cite{petrenko} 
the protons are actually responsible for the
amazing electrical properties of ice \cite{hobbs}. The current
understanding of the motion of protons in H-bond materials is mainly
based in a few phenomenological models. In this paper we discuss
a qualitative microscopic model which captures the basic quantum
mechanical correlations of the proton system.

We show that
the physics of protons in ice is a clear example of a strongly
correlated problem very similar to the ones discussed in strongly
correlated electron systems \cite{Fradkinbook} reminiscent of the
physics of Mott insulators \cite{mott}. More specifically, the
motion of protons in ice is hindered by strong constraints
that forbid single proton hopping and only allows for collective
ring-exchange-like motion. This sort of systems are known to be
closely related to the physics of gauge theories.
As we will see below, in the phase of ice in which the protons are ordered, 
in analogy with the problem of confinement of
quarks in hadronic matter, pairs of defects (anions and cations)
cost an energy which is linear with the separation between them.

Although there is a vast literature on physics of ice \cite{hobbs}, 
we focus only on the quantum motion of protons. The concepts introduced here
can be easily extended to the study of many different problems in H-bond materials.
Ice shows many different solid phases depending on temperature and pressure.
These phases are essentially controlled by the geometry of the
H-bonds that link different water molecules.
Furthermore, a
H atom is located asymmetrically relative to the two
O atoms, forming a double well structure in the bond. Thus,
a H atom can sit in any of the two sides a bond. 
Hence, we can think
of ice as made out of protons, H$^+$, moving in a crystal lattice
made out of O$^{-2}$ ions. 
Therefore, solids made out of H-bonds are expected to have strong
electrical properties. For example, ice exhibits a high static
permittivity comparable with
the one of liquid water, electrical mobility
that is large when compared to most ionic conductors (in fact,
the mobility is comparable to the electronic conduction in metals). 
These are striking properties
since in the solid phase only protons can move in an ice lattice 
as the electrons form a band insulator \cite{petrenko}.

A striking feature of ice is its extensive classical entropy
at low temperatures \cite{hobbs}.
This large entropy implies a macroscopic number of classically 
degenerate states at low temperatures. This is a situation very similar to frustrated 
magnetic systems, which the best representative is precisely called 
spin ice model \cite{Harris,gingras}. 
Our study,
however, focus entirely on the proton motion in ice and not on 
magnetism. As we show, the proton motion in ice can be
mapped into a problem with pseudo-spins, in close analogy to
some frustrated magnets.

The most successful
explanation for the physical behavior of ice was given by the phenomenological
work of Bernal and Fowler \cite{bf} in 1933 that gave rise to the so-called
Bernal-Fowler (BF) rules:
({\it i}) the orientation of H$_2$O molecules is such that only
one H atom lies between each pair of O atoms; ({\it ii})
each O atom has two H atoms closer to it forming a water molecule.
Rule ({\it i}) prevents situations in which a H-bond has two
H atoms. Rule ({\it ii}) does not
allow for configurations in which each O has more than two
H close to it.
As it was shown by Pauling \cite{pauling}, using a classical counting
argument, these rules can
account for most of the entropy measured in the experiments, and
predict an extensive entropy even at $T=0$.
A more detailed calculation taking into account ring exchange of protons
in ice crystal brings this number even closer to the measured
value \cite{nagle}. Pauling's statistical model has been very
successful in explaining the distribution of protons in ice and
has been confirmed experimentally in nuclear magnetic resonance (NMR) 
and neutron scattering experiments \cite{NMR}.

Although Pauling's calculation explains the
arrangement of protons in ice it fails to describe its
electrical properties. The reason for this failure is due to
the fact that any motion of
the protons under the BF rules requires an
extremely correlated behavior \cite{ihm}. In order to explain
the electrical behavior another phenomenological concept was introduced,
namely, the concept of {\it defects}. Defects, by definition, are
local violations of the BF rules. The defects associated
with violation of rule ({\it i}) are called Bjerrum defects \cite{bjerrum}
and the ones associated with 
violations of rule ({\it ii}) are called ionization
defects \cite{defect}. With the concept of defects
one is able to explain most of the electrical properties of ice \cite{hobbs}.

Protons can move by the rigid rotation of the water molecule, by thermal
activation over an energy barrier
from one side of the H-bond to another, or by {\it quantum
tunneling} under the energy barrier between the two sides of the bond.
At low temperatures, the rigid motion of the molecule and thermal
activation are exponentially suppressed and only quantum tunneling
is allowed. Quantum tunneling is possible because the wavefunction
of the proton is extended from one side to another in the bond.
Although the phenomenological theories of ice can
account for a great part of the experimental data
there are still many experiments that remain unexplained such as
anomalies in the specific heat in pure \cite{giauque}
and doped ice \cite{pick}. Originally it was proposed by
Onsager \cite{onsager} that these anomalies could be explained
by a ferroelectric transition. It turns out, however, that
there is no evidence for any polar effect in ice. Quite to the
contrary, the BF rules state that ice is a rather non-polar material.
In this paper we show that by a quantum mechanical mechanism of
order-by-disorder, an ordered phase of protons emerges at low temperatures.
The melting of this ordered state can account for the specific heat
anomalies.

The classical ice model is defined on a pyrochlore lattice and its
planar representation is the checkerboard lattice. The planar
model is equivalent to the six vertex model what has been
solved exactly by Lieb \cite{Lieb}. It is known from these studies
that the phase diagram of the classical planar system has two
phases, an ordered anti-ferroelectric phase and a line of critical
points with extensive entropy at zero temperature. The anti-ferroelectric
phase corresponds to a staggered arrangement of proton positions in
the H-bonds. Its quantum version has been
studied by Moessner, Sondhi \cite{MS} and collaborators
\cite{others,olav}. 

In this paper, we consider the quantum version of the ice problem
and show that quantum fluctuations stabilize the anti-ferroelectric
phase in the absence of defects by a mechanism of order-by-disorder.
To this end, we use a 
gauge theory description of the planar quantum ice problem which
can be easily extended to higher dimensional systems.  This gauge
theory has also being used in similar models such as
the quantum dimer models on the square lattice \cite{Fradkinbook}. 

This paper is organized as follows: In the next section we discuss 
the characteristic energy scales of the ice problem and propose
the minimal model for proton motion in ice. Here we discuss in detail
the analogy and connection to frustrated quantum magnets. 
Section \ref{mapping} contains
the mapping of the planar ice model to a gauge theory.
In Section \ref{groundstate} 
we determine the ground state of the
problem in the neutral sector. Here we show how an ordered state of protons
arises from the order-by-disorder mechanism and how
the ionic defects are confined in the planar model. 
In Section \ref{threedimensions} we discuss the gauge problem
in three dimensions and obtain the phase diagram including thermal effects. 
We show that the three dimensional case has a confining-deconfining
transition even at T=0.  In Section \ref{cphonons} we briefly discuss the 
effects
of lattice vibrations on the proton motion. We argue that 
the second order phase transitions obtained in the case without
phonons can become first order. Section \ref{conclusions} contains our 
conclusions and the comparison between the theory and the experimental
measurements in hexagonal (Ih) ice.

\section{The model}
\label{model}

\begin{figure}[htb]
\centerline{\includegraphics[width=6cm,keepaspectratio]{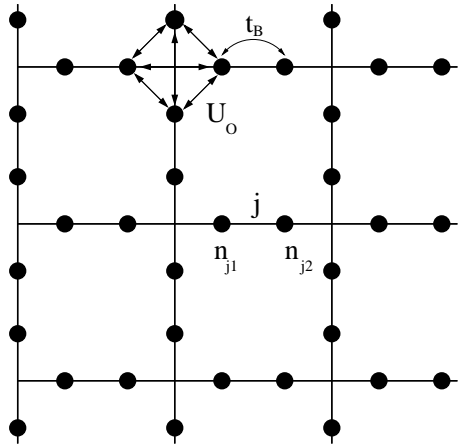}}
\caption{Planar ice model: the vertices are O sites and the dots are H
  sites. The symbols are explained in the text.}
\label{square_ice}
\end{figure}

Consider a lattice with protons living in the bonds and where
each vertex is to be interpreted as an O atom as shown in
Fig.\ref{square_ice}.
The lattices we are going o discuss in this papers are the 
planar square lattice,
the cubic lattice and the pyrochlore lattice.
The protons can occupy two positions in their respective link.  
Let us divide any of those lattices into sub-lattices 1 and 2, and for each
link $i$ define 
the sites $i1$ and  $i2$ neighboring
a vertex of sub-lattice 1 and 2, respectively. 
We can define the proton occupation
number $n_{i,a}$, $a=1,2$ for each proton site $ia$ of the system. 

The main energy scales in this problem come from the Coulomb repulsion 
between the protons: the on-site Coulomb repulsion, $U_H$ 
(the so-called Hubbard term), the Coulomb repulsion between protons 
in the same H-bond, $U_B$, and the
Coulomb repulsion between protons around the same O vertex, $U_O$ \cite{lekner}. 
The physical situation of interest for ice corresponds to the case where 
$U_H \gg U_B \gg U_O$.  Notice that this condition ensures that there is 
only one proton per lattice site, only one proton per bond, and for $1/4$ 
filled ice it implies 
that there are only two protons around the O vertex. Thus, the
electrostatic repulsion between the protons leads to the ice rules. 

Since we have considered interactions up to second nearest neighbors one
might wonder whether we should not include interactions with even longer
range. In fact, it is well-known that the water molecules interact through
long-range dipole forces that in our picture would be represented by 
long-range interactions between H$^+$ ions. The same problem arises in
the context of spin-ice \cite{gingras,gingraslongrange}. Monte carlo
simulations \cite{montecarlo} for the frustrated lattices discussed here have shown that 
the long-range dipolar interactions "average out" to zero at large distances. 
This result shows that the ice rules are still obeyed in the absence of 
long-range interactions and that the degeneracy of the classical state
is not lifted. Moreover, improved Monte Carlo simulations with "non-local"
updates detected the appearance of long range order only at very low energies,
much smaller than the energy scales discussed here \cite{nonlocal}. Thus,
from now on the effects of long-range interactions between second nearest 
neighbors will be disregarded.

The Bjerrum and ionic defects mentioned earlier only occur if the protons
hop from site to site. There are essentially only two types of hopping
in this lattice: hopping on the H-bond with energy $t_B$ and hopping across
the O vertex, $t_O$. Because the hopping energy is an exponential function
of the distance it is easy to show that $t_B \gg t_O$. 
The presence of $t_B$ allows for the hopping of
the protons around the O vertices leading to the creation of ionic defects.
The hopping $t_O$, on the other hand, allows for two protons in the same
H-bond and therefore can lead to Bjerrum defects. The values of $t_B$ and
$t_O$ are much smaller than in electronic systems because of the much larger
proton mass. Thus, we expect that $U_O \gg t_B$ making the proton
motion analogous to the motion of electrons in Mott insulators.   

\begin{figure}[htb]
\centerline{\includegraphics[width=6cm,keepaspectratio]{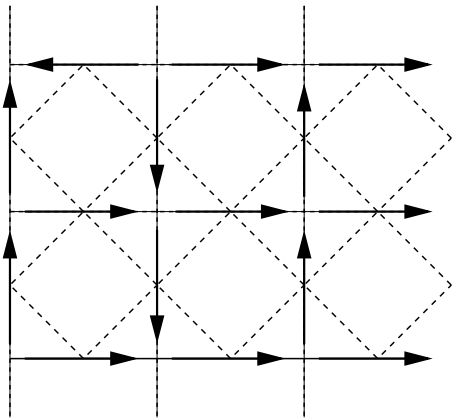}}
\caption{A possible configuration of the pseudospins that satisfies the BF rules.}
\label{isingconf}
\end{figure}

In what follows we assume that $U_H \gg U_B \to \infty$ and $t_B \gg t_O \to 0$
so that only ionic defects are possible in the problem. It is also clear that
in this limit the classical ground state obeys the ice rules and that
magnetic phenomena associated with the proton spin does not play any role.
Within the assumption that the on-site Coulomb repulsion between protons is
very large, we restrain to the case
where $ n_{ia} =0,1$, $\forall i$ and the proton spin degrees of freedom 
can be ignored.
The Hamiltonian for the proton motion on neutral planar ice with
just one proton per bond, that is,
\begin{eqnarray}
n_{i1}+n_{i2} = 1 \, ,
\label{constraint}
\end{eqnarray}
is written as, 
\begin{eqnarray}
{\cal H} &=& U_O \sum_{\langle i,j \rangle} \left(n_{i1} n_{j1} +
  n_{i2}n_{j2}\right)
\nonumber
\\
&-& t_B \sum_i \left(c^{\dag}_{i1} c_{i2} + h.c. \right) \, .
\label{hamileff}
\end{eqnarray}
We now make use of the constraint
Eq.\eqref{constraint}, and define pseudo-spin operators associated to link $i$
as (we use units such that $k_B = 1 = \hbar$):
\begin{eqnarray}
S_{i}^x &=& \frac{1}{2} \left(c^{\dag}_{i1} c_{i2} + h.c.\right) \, ,
\nonumber
\\
S_i^y &=& \frac{i}{2} \left(c^{\dag}_{i1} c_{i2} - h.c.\right) \, ,
\nonumber
\\
S_i^z &=& \frac{1}{2} \left(c^{\dag}_{i1} c_{i1} - c^{\dag}_{i2}
  c_{i2}\right) \, ,
\label{transformation}
\end{eqnarray}
which, due to condition Eq.\eqref{constraint},
obey the spin algebra $[S_{i}^x,S_j^y] = i \delta_{i,j} S_i^z$.
In terms of these operators the Hamiltonian  of Eq.\eqref{hamileff}
is written as
\begin{eqnarray}
{\cal H} =  J \sum_{\langle i,j \rangle} S_i^z S_j^z -
\Gamma \sum_i S_{i}^x \, ,
\label{ising}
\end{eqnarray}
where $J = 2 U_O$ and $\Gamma = 2 t_B$. In the limit of $\Gamma=0$ the ground 
state is highly degenerate because there are many configurations of the 
pseudospins that give the same energy. One possible configuration is shown
in Fig.\ref{isingconf}.

\begin{figure}[htb]
\centerline{\includegraphics[width=6cm,keepaspectratio]{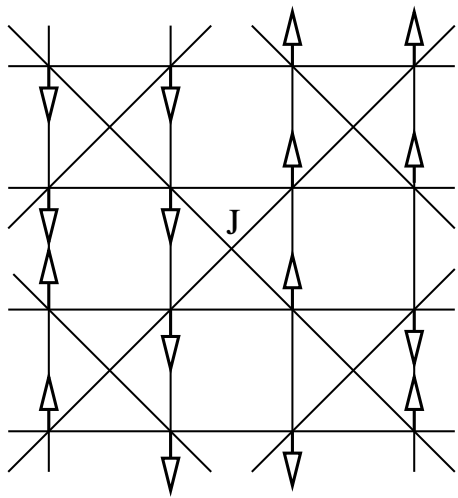}}
\caption{Effective spin lattice for the ice problem in the Mott regime. The
  pseudospins interact with an antiferromagnetic exchange $J$. }
\label{checkerboard}
\end{figure}

By a duality transformation, in which each link corresponds to a 
vertex of the square dual lattice, it is easy to show that Eq.\eqref{ising} describes
also the Ising model in a transverse field on a lattice where now
the spins are defined on vertices. 
In the case of the planar ice system, this lattice correspond
to the checkerboard lattice (see Fig.\ref{checkerboard}), while for the
pyrochlore and cubic lattices one obtains corner sharing tetrahedra and hexades, 
respectively.
It is clear that the ground state of the Hamiltonian of Eq.\eqref{ising}
with $\Gamma = 0$ is an eigenstate of the operator $S^z$.
This state corresponds the
proton localized everywhere in either side of the link.
In order to minimize the energy the
distribution of protons is such that the only contributing
configurations have the total magnetization of each vertex is equal to zero.
For the pyrochlore and planar lattices, 
this means that each O has two protons close to
it and two away. For the cubic lattice one would have three protons close to
the O and three away.
The number of configurations that satisfy the ice rules grows exponentially 
with the size of the
system. Taking the planar case as an example, there are
$\left(4/3\right)^{3N/2}$ of them~\cite{Lieb}. 

\begin{figure}[htb]
\begin{center}
\subfigure[]{\includegraphics[width=0.4\textwidth]{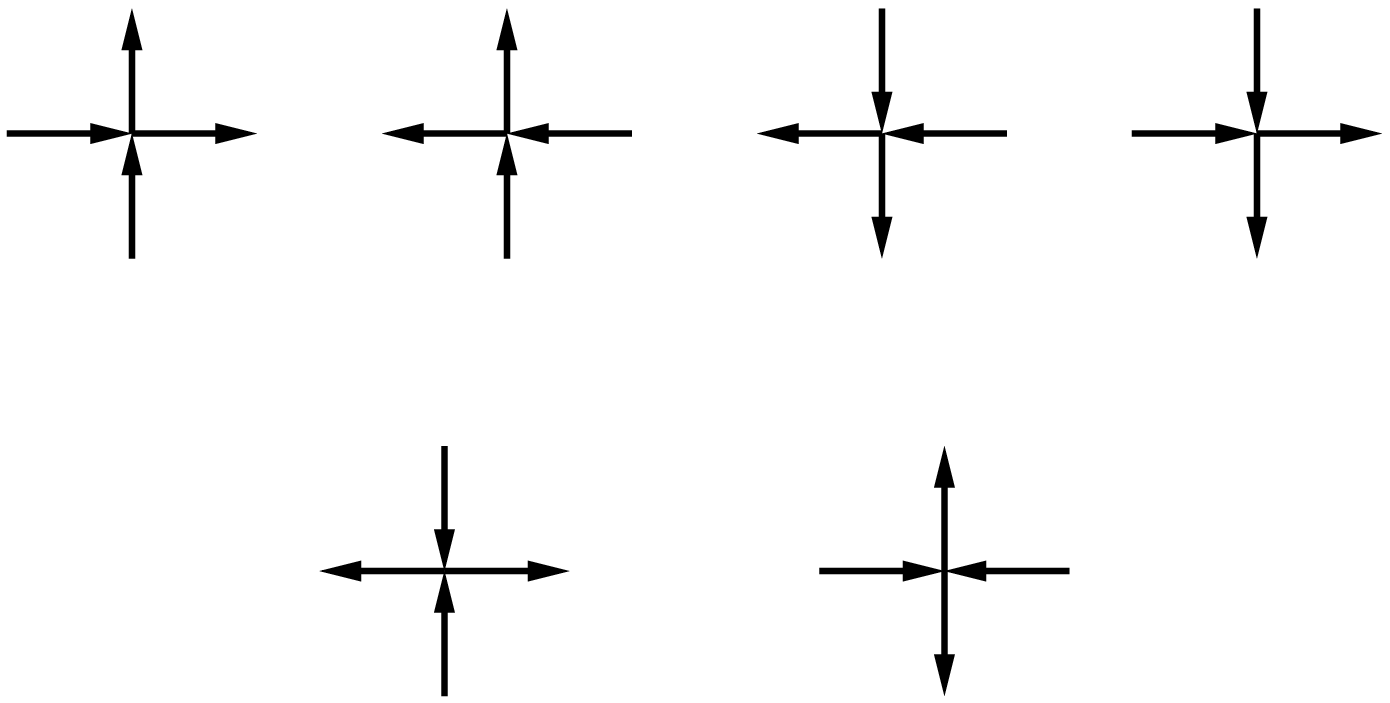}}
\\
\subfigure[]{\includegraphics[width=0.4\textwidth]{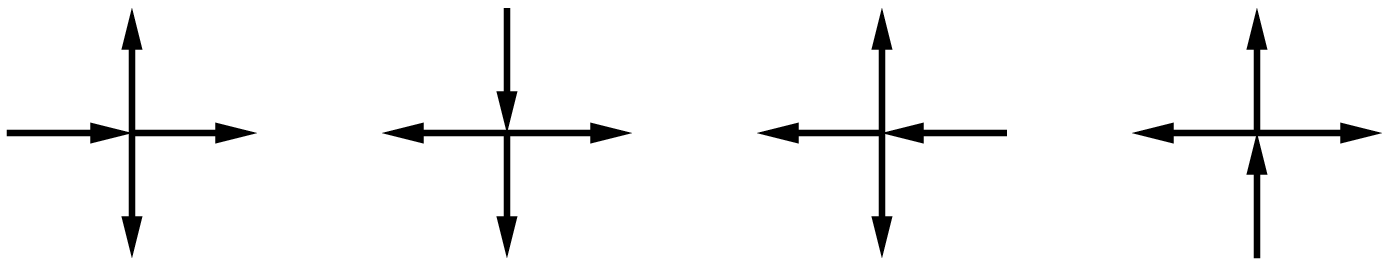}}
\\
\subfigure[]{\includegraphics[width=0.4\textwidth]{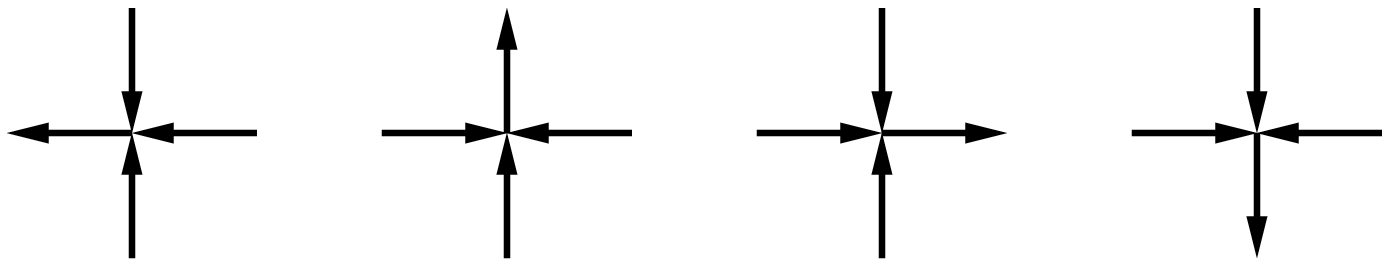}}
\end{center}
\caption{(a) Configurations of the pseudospins that obey the BF rules;
(b) ionic defect with charge $-$ (OH$^-$); (c) ionic defect with charge
 $+$ (H$_3$O$^+$).}
\label{configurations}
\end{figure}

In the checkerboard language (see Fig.\ref{checkerboard}) a sublattice of
plaquettes contains next nearest neighbor interactions (``crossed plaquettes'').  
According to the BF rules the plaquettes have total
$S^z$ magnetization equal to $0$ as shown in Fig.\ref{configurations}(a).
The configurations that contain defects violate the ice rules as shown in
Figs.\ref{configurations}(b) and (c). 
Classically, as it is well known, the system without defects corresponds
to the six vertex model (see Fig.\ref{configurations}(a)). 
The full classical problem is a 14 vertex
model that, as far as we know, has not been solved exactly \cite{baxter}.

Let us pick, for example, a particular configuration of the system that
satisfies the ice rule and let us flip a single proton. This creates a
pair of $+$ (H$_3$O$^+$) and $-$ (OH$^-$) ionic charges at 
two neighboring O sites, corresponding to the configurations
Fig.\ref{configurations}(c) and Fig.\ref{configurations}(b), respectively. 
Strictly at $\Gamma = 0$, one can separate
the $+$ and $-$ defects along a zig-zag-like trajectory at
no extra energy cost. There is however an entropic price for separating
such defects that introduces an effective interaction between
them. These charge defects play the role similar to holons in quantum
dimer models \cite{Fradkinbook}. Two natural questions arise in this
context. The first one concerns possible lifting of the degeneracy
of the ground-states by quantum fluctuations when $t_B$ or $\Gamma$
are finite. The second question is whether in the presence of
these quantum fluctuations the ionic defects can be separately
freely (unconfined) as in the classical case.

To answer the first question, let us consider the operator
$P$ that projects into the subspace of states that
satisfies the ice rules. Let $PHP^{-1}$ denote the Hamiltonian of 
Eq.\eqref{ising} projected onto the ice rules sector, and let us
treat the effects of finite $\Gamma$ in perturbation theory.
The effective Hamiltonian, for the planar and cubic lattices, 
to the lowest non-vanishing order is (${\bf S}=\vec{\sigma}/2$):
\beq
H_{eff} = -t \sum_{\square} \left( \sigma_{ij}^+ \sigma_{jk}^- \sigma_{kl}^+
 \sigma_{li}^-  + h.c. \right) \, ,
\label{heff}
\eeq
where $ij$ denote the links belonging to a crossed plaquette. To lowest order
we find $t~\sim~\Gamma^4/J^3$.
For the pyrochlore lattice, the lowest order effective Hamiltonian is:
\beq
H_{eff} = -t \sum_{ijklmn} \left( \sigma_{ij}^+ \sigma_{jk}^- \sigma_{kl}^+
 \sigma_{lm}^-   \sigma_{mn}^+ \sigma_{ni}^- + h.c. \right) \, ,
\label{heff2}
\eeq
where now the interaction is around an hexagonal plaquette corresponding to
the smallest loop in the pyrochlore lattice, and $t~\propto~\Gamma^6/J^5$.
The planar lattice with the Hamiltonian of Eq.\eqref{heff} has been studied at zero
temperature $T=0$~\cite{MS}. It can be
mapped onto a height model in which quantum fluctuations select the ordered
flat state corresponding to a N\'eel order in the checker-board model.
It is worth mentioning that the effective Hamiltonian of Eq.\eqref{heff} and
Eq.\eqref{heff2} also describe
tunneling within the low energy  manifold of the $XXZ$ spin $1/2$ model
on the checkerboard.
To see that, we start by writing the $XXZ$ Hamiltonian:
\beq
H = \sum_{<ij>} \left(J_z S^z_i S^z_j + {J_{xy} \over 2} ( S^+_i S^-_j + S^-_i
  S^+_j ) \right) \, ,
\label{HXXZ}
\eeq
where $<ij>$ stand for all the couple of links belonging to the checkerboard lattice.
We assume $J_{xy} \ll J_z$.
At zeroth order in $J_{xy}$ the ground state manifold is identical to
the one mentioned before, with total $S^z$ magnetization per crossed plaquette equal to zero.
Considering then the $XY$ term in the first non-vanishing order in perturbation
theory leads to Eq.\eqref{heff} with $t~\propto~J_{xy}^2/J_z$ and Eq.\eqref{heff2} with 
$t~\propto~J_{xy}^3/J_z^2$, respectively. 

\section{Mapping of the planar quantum model to a lattice gauge theory}
\label{mapping}

In this section we map the ice problem onto a gauge theory \cite{kogut,fradkin}.
Consider the O lattice which is shown schematically in Fig.\ref{square_ice}.
The H-bonds form links between the O atoms and to each link
we can assign a value given by the pseudo spin defined in the
previous section. Associated with these links we can assign states
corresponding to the two configurations of the protons in each link
which we will denote as $|+1/2\rangle$ and $|-1/2\rangle$.
The ice rules imply
\beq
\sum_{i \in \square} \sigma^z_i = 0 \, ,
\eeq
where $\sigma_z |\pm 1/2 \rangle = \pm |\pm 1/2 \rangle$. 
This Ising gauge theory is equivalent to the one
considered in Ref. [\onlinecite{MSF}], and as in that case, it has a local 
$U(1)$ symmetry generated by the unitary operator
\beq
{\cal U}_v = \prod_{\alpha_v} e^{i \alpha_v
  \sum_{i \in v} \sigma^z_i} \, ,
\eeq
with each $\alpha_v$ associated to the vertex $v$ being arbitrary.
This approach has been applied with success in the case of the
quantum dimer model \cite{MSF}. 

There is an alternative and complementary approach which works in
an enhanced Hilbert space and that was applied originally to
the quantum dimer model as well \cite{Fradkinbook,steve}.
Let us consider a square lattice is spanned by vectors
${\bf r}=(n {\bf e}_1 +m {\bf e}_2 )a$
where $n$ and $m$ are integers, ${\bf e}_1={\bf x}$,
${\bf e}_2={\bf y}$ and $a$ is the lattice spacing.
On each link (i.e., H-bond) we define a variable $\tilde{\ell}_{\mu}({\bf r})$
with $\mu=1,2$ which can be $-1/2$ or $1/2$, which transfer like a vector:
$\tilde{\ell}_{-\mu}({\bf r}+{\bf e}_{\mu}) = - \tilde{\ell}_{\mu}({\bf r})$.
It will be convenient to define a new variable  $\ell_{\mu}
= \tilde{\ell}_{\mu} +1/2$ which can take values of $0$ or $1$.
Let us denote by $\Delta_{\mu}$ the discrete derivative,
\begin{eqnarray}
\Delta_{\mu} \ell_{\nu}({\bf r}) = \ell_{\nu}({\bf r}+{\bf e}_{\mu})
- \ell_{\nu}({\bf r}) \, .
\end{eqnarray}
It is easy to see that the ice rules are equivalent to impose the
condition that
\begin{eqnarray}
\sum_{\mu=1,2} \Delta_{\mu} \ell_{\mu}({\bf r}) = 0 \, .
\label{gauss}
\end{eqnarray}
Only configurations of the form Fig.\ref{configurations}(a)
are allowed for $\tilde{\ell}_{\mu}$. 
Therefore, Eq.(\eqref{gauss} reflects the BF rules. 

Eq.\eqref{gauss} has the form of Gauss law in electrodynamics without
external charges, $\nabla \cdot {\bf E} =0$.
We will then assign to violations of the ice rules, such as those in 
Fig.\ref{configurations}(c) the condition of 
$\sum_{\mu=1,2} \Delta_{\mu} \ell_{\mu}({\bf r}) = + 1$
for the formation of a H$^+_3$O and 
$\sum_{\mu=1,2} \Delta_{\mu} \ell_{\mu}({\bf r}) = - 1$
for the formation of HO$^-$.
Thus, if we allow violations of the ice rules by formation of
ionization ``defects" we must have
$\sum_{\mu=1,2} \Delta_{\mu} \ell_{\mu}({\bf r}) = Q$
where $Q$ plays the role of the ``effective'' charge 
of the defect. 
In the quantum theory we define a Hilbert space of states 
$|\ell_{\mu}({\bf r})\rangle$ which are the eigenstates of
the operator $E_{\mu}({\bf r})$:
\begin{eqnarray}
E_{\mu}({\bf r}) |\ell_{\mu}({\bf r})\rangle =
\ell_{\mu}({\bf r}) |\ell_{\mu}({\bf r})\rangle \, .
\end{eqnarray}
In the quantum theory Gauss law is a constraint in the
space of state: 
\begin{eqnarray}
\sum_{\mu=1,2} \Delta_{\mu} E_{\mu}({\bf r}) |{\rm Phys.}\rangle = 0 \, ,
\label{gaussquantum}
\end{eqnarray}
which defines the physical states of the system.
Let $\theta_{\nu}$ be the operator canonically conjugated to $E_{\mu}$, 
\begin{eqnarray}
\left[\theta_{\nu}({\bf r'}),E_{\mu}({\bf r})\right]= i \delta_{{\bf r},{\bf r'}}
\delta_{\mu,\nu} \, .
\end{eqnarray}
Thus, we see that $E_{\mu}$ plays the role of the electric field while
$\theta_{\nu}$ plays the role of the vector potential in electrodynamics.

Let us now define a ring exchange operator such that it maps configuration
of electric fields satisfying the ice rule conditions. In this context
this amounts to require that these operators, when acting on a plaquette,
change the configuration in a manner consistent with Eq.\eqref{gaussquantum}.
For each plaquette $p$ we define the operator $e^{i \Phi(p)}$
where
\begin{eqnarray}
\Phi(p) = \sum_{{\bf r},\mu \in p} \theta_{\mu}({\bf r}) \, .
\end{eqnarray}
This operator is gauge invariant in the sense that it commutes with
the generator of gauge transformations, $\Delta_{\mu} E_{\mu}({\bf r})$.
When acting on the links one gets, 
\begin{eqnarray}
e^{\pm i \Phi(p)} |\ell_{\mu}({\bf r})\rangle = |\ell_{\mu}({\bf r})\pm
1\rangle \, .
\label{flip}
\end{eqnarray}

We are now going to relax the constraint that
$\ell_{\mu}({\bf r})$ can be $0$ or $1$ by enlarging the Hilbert
space to all integer values of $\ell_{\mu}({\bf r})$.
However, we will penalize energetically the values of $\ell_{\mu}({\bf r})$ 
that are not $0$ or $1$ by adding an extra term to the Hamiltonian
of the form:
\begin{eqnarray}
H_0 = \frac{1}{2 g} \sum_{{\bf r},\mu}
\left[\left(E_{\mu}({\bf r})-\frac{1}{2}\right)^2 - \frac{1}{4}\right] \, .
\label{lagrange}
\end{eqnarray}
It is obvious that when $g \to 0$ $\ell_{\mu}({\bf r})$ can only be $0$ or $1$.
Together with the constraint of Eq.\eqref{gaussquantum}, Eq.\eqref{lagrange} defines the classical problem and
the BF rules. Note that, for example, the N\'eel state corresponds to a
configuration of electric fields in which the links (horizontal and
vertical) form a staggered configuration of flux 0 and 1. These fluxes form
stair like lines (right-up-right-up...) winding around the system.
We also note that if the $1/2$ term in the kinetic energy Eq.\eqref{lagrange}
where not present,
this problem would be equivalent to compact electrodynamics with matter
fields studied in Refs. [\onlinecite{kogut,fradkin}].
The quantum kinetic energy of this problem is given by: 
\begin{eqnarray}
H_I = - t \sum_p \cos \Phi(p) \, ,
\label{quantum}
\end{eqnarray}
which is analogous to the magnetic energy in compact electrodynamics.

In order to introduce ionic
``defects" one has to introduce the charge $Q$ into the problem.
Associated with $Q$ one defines the quantum operator $n({\bf r})$,
which counts the number of defects at each vertex, and its conjugate $\phi({\bf r})$
such that
\begin{eqnarray}
\left[\phi({\bf r'}),n({\bf r})\right] =  i \delta_{{\bf r'},{\bf r}} \, .
\end{eqnarray}
where $n({\bf r})=0,\pm 1, \pm 2,...$
The Hamiltonian for the ``matter" 
field $\phi$ contains a term for the energy required to create any one of these charges.
For instance, one could write,
\begin{eqnarray}
H_Q = E_0 \sum_{\bf r} n^2({\bf r}) \, ,
\label{hq}
\end{eqnarray}
where $E_0$ is the energy required to create a OH$^{-}$-H$^+_3$O pair.
Observe that Eq.\eqref{hq} tends to suppress defects.
The motion of the defects is given by a kinetic energy term which is
\begin{eqnarray}
H_C = \lambda \sum_{{\bf r},\mu} \cos \left(\Delta_{\mu}\phi({\bf r})
 -  \theta_{\mu}({\bf r})\right) \, ,
\label{coup}
\end{eqnarray}
where $\lambda \sim \Gamma$ is the coupling constant. 
This kinetic energy is gauge invariant since it commutes with the
generator of local gauge transformations 
\begin{eqnarray}
U = e^{i \sum_{{\bf r}} \alpha({\bf r}) \hat{G}({\bf r})} \, ,
\end{eqnarray}
where
\begin{eqnarray}
\hat{G}({\bf r}) = \sum_{\mu=1,2} \Delta_{\mu} E_{\mu}({\bf r})-
n({\bf r}) \, .
\end{eqnarray}

The gauge theory of the ice problem is 
described by the Hamiltonian $H = H_0+H_I+H_Q+H_C$
defined by Eqs.~\eqref{lagrange}, \eqref{quantum}, \eqref{hq}, 
and \eqref{coup}. 
In the next section we are going to study the ground state of
such a theory in the neutral sector \cite{fendley}.

\section{Ground-state selection in the neutral sector and confinement}
\label{groundstate}

\begin{figure}[htb]
\centerline{\includegraphics[width=6cm,keepaspectratio]{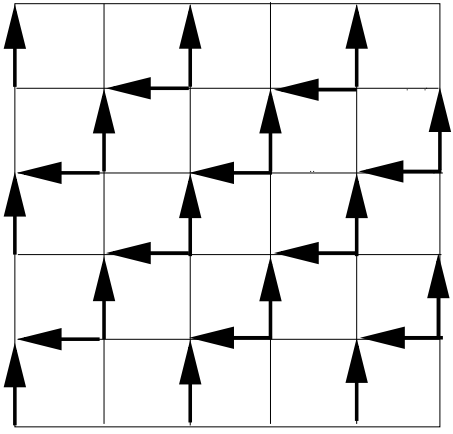}}
\caption{Background field $B^N_{\mu}$.}
\label{pierre}
\end{figure}

It is interesting to consider the analysis of the $U(1)$ lattice
gauge theory of quantum dimer models given in 
Refs.~[\onlinecite{Fradkinbook,steve}]
in the context of our calculation.
In the neutral sector, since the electric field has zero
divergence, it can be written as:
\beq
E_{\mu}{\bf r}) = \epsilon_{\mu \nu} \left[ \Delta_{\nu} S({\bf r})
+ B_{\nu} ({\bf r}) \right] \, ,
\label{e=s+b}
\eeq
where $S$ is an integer valued function defined on the dual lattice
with periodic boundary conditions
and $B_{\nu}$ is defined on the links of the dual lattice.
Since the Gauss law tells us that
$$
\epsilon_{\mu \nu}  \Delta_{\mu} B_{\nu} = 0 \, ,
$$
only non-trivial topological configurations of $B_{\nu}$ have to be
considered. Let us choose, in particular,  $B^N_{\nu}$ constructed
in the following way:
let us define $B^N_{\mu}$ for $\mu=2$ to take the alternating values $0$ and
$1$ on vertical links from row to row. For horizontal links with $\mu=1$, 
$B^N_{\mu}$ will alternate from $0$ and $-1$ in such a way that each positive oriented arrow in
the vertical direction meets a negative oriented arrow in the horizontal
direction (see Fig.\ref{pierre}). By choosing such a $B^N_{\nu}$
and choosing $S=0$ on the dual sites one gets for $E_{\mu}$ one of the two
N\'eel configuration (the other configuration being obtained by assigning 
a $0,1$ staggered value to $S$).

We can now write the path integral representation in discrete imaginary time
in the form of the a 3-D discrete Gaussian model:
\begin{eqnarray}
{\cal S} &=& {1 \over 2t \epsilon} \sum_{{\bf r}, j} \left[ \Delta_0 S({\bf r},j)
  \right]^2 \nonumber
\\ &+& {\epsilon  \over 2g} \sum_{{\bf r}, j, \mu = 1,2} \left[
\Delta_{\mu} \left( S({\bf r},j) - S^0 \right) \right]^2 \, ,
\end{eqnarray}
where $j$ is the discrete time coordinate ($\epsilon$ and $\Delta_0$ are the
``lattice spacing" and the discrete derivative in the imaginary time direction, respectively), and
$S^0$ is a staggered configuration alternating between $0$ (in say,
sub-lattice ``A'') and $1/2$ (in sub-lattice ``B''). While a non-zero
$t$ allows for fluctuations of the field $S$, only some kind of fluctuations
can be made without paying the energy of the huge ``surface tension'' $1/g$.
More precisely, if at some point of sub-lattice ``A'' the value of $S$ is 0,
then the neighboring values of $S$ (belonging to sub-lattice ``B'') can
fluctuate between the values $-1$ and $0$ with no cost in surface tension.

Imagine now that we consider the most general situation for the electric field
Eq.\eqref{e=s+b} in which the vector field $B_\mu$ has a non-trivial winding.
The action for the discrete Gaussian model becomes :
\bea
{\cal S} &=& {1 \over 2t \epsilon} \sum_{{\bf r}, j} \left[ \Delta_0 S({\bf r},j)
  \right]^2 \nn \\
  &+& {\epsilon  \over 2g} \sum_{{\bf r}, j, \mu = 1,2} \left[
\Delta_{\mu} \left( S({\bf r},j) - S^0 \right) -B_\mu \right]^2 \, .
\eea
Let us also assume that we choose the configuration for $B_\mu$
with the minimal number of links where it is non-zero.
One can show that, for any configuration with non-zero $B_\mu$, either there is
no configuration of the $S$ degrees of freedom that minimizes the
surface tension term at each point, or at each point connected to a link
where $B_{\mu}$ is non-zero, there is a unique value for the
field $S({\bf r},j)$ (modulo a global constant) that satisfies both periodic
boundary conditions and a minimal surface tension. In the limit $g \rightarrow
0$ the values of the field $S$ associated to those points are not allowed to
fluctuate.  Then, the computation of the partition function of a discrete
Gaussian model in a topological sector having non-zero $B_{\mu}$ is penalized by
the surface tension term or gives rise to the
freezing of some of the plaquette degrees of freedom $S({\bf r},j)$ which gives
a smaller contribution than the sector $B_{\mu}=0$ (since the configuration space of
the former is a subset of the later).
Keeping in mind these arguments, from now on, we are going
to consider only the sector $B_{\mu}=0$.

To deal with the discrete variable
$S$ we make use of the Poisson summation formula:
\beq
\sum_{n = -\infty}^{\infty} f(n) = \sum_{m = -\infty}^{\infty} \int ~ d\phi
~ e^{i 2 \pi m \phi } f(\phi) \, ,
\eeq
and write the partition function:
\bea
Z \sim \sum_{\{m\}} \int ~ D\phi ~ \exp ({\cal S}(m,\phi)) \, ,
\eea
where, 
\bea
{\cal S} &=& \sum_{{\bf r}, j} \left\{ i 2 \pi m({\bf r},j)
  \phi({\bf r},j) - {1 \over 2t \epsilon} \left[ \Delta_0 \phi({\bf r},j)\right]^2  \right.  
  \nonumber \\
&-& \left. {\epsilon  \over 2g} \sum_{\mu = 1,2} \left[
\Delta_{\mu} \left( \phi({\bf r},j) - S^0 \right) \right]^2 \right\} \, .
\eea
We redefine: $\tilde{\phi} = \phi - S^0$, and write
$\tilde{\phi}$ as $\tilde{\phi}_A$ and $\tilde{\phi}_B$ for sub-lattices A
and B, respectively. Following Ref.~[\onlinecite{RS}], we work on the dilute
gas approximation and  keep only
``monopoles'' of charge $\pm 1$. In this case one finds an effective action of the form:
\bea
{\cal S} &=& \sum_{{\bf r}, j} \left\{ {\epsilon  \over 2g} \left(\tilde{\phi}_A -
\tilde{\phi}_B \right)^2
+ {1 \over 2t \epsilon} \left(\left[ \Delta_0
\tilde{\phi}_A \right]^2  + \left[ \Delta_0 \tilde{\phi}_B \right]^2 \right) 
\right. \nonumber \\
&-& \left. z ~ \cos(2 \pi \tilde{\phi}_A) + z ~ \cos(2 \pi \tilde{\phi}_B)  \right\}\, ,
\label{tots}
\eea 
where $z$ is the monopole fugacity.
In order to take the continuum limit, we write
\beq
\tilde{\phi}_A = \chi_1 + \chi_2 ~~,~~ \tilde{\phi}_B = \chi_1 - \chi_2 \, ,
\eeq
Note that a non-zero expectation value for $\chi_2$ corresponds
to the two N\'eel orders depending on the sign of $<\chi_2>$ while
$<\chi_2>=0$ corresponds to a disordered state.
Expanding in derivatives of the fields we get a continuum action of
the form:
\bea
{\cal S} &=& \int d^2\vec{r} d\tau \left\{ \frac{1}{2} \left[ K_x (\nabla \chi_1)^2 + K_{\tau}
  (\partial_{\tau} \chi_1)^2  + \alpha~ \chi_2^2\right] \right. \nonumber \\
 &+& \left. \lambda ~ \sin(2 \pi \chi_1)
\sin(2 \pi \chi_2) \right\}  \label{effactionfreezed} \, .
 \eea 
Where $\nabla$ is the spatial two-dimensional gradient, $K_x$, $K_{\tau}$, $\alpha$,
and $\lambda$ are the parameters of the effective coarse grained model that in principle, but with a
large amount of effort, can be calculated from the microscopic theory. 
Here the continuum limit has being taken with the prescriptions: 
\bea
\Delta_0 \chi({\bf r},j) &\rightarrow& \epsilon  \partial_{\tau}  \chi({\bf r},\tau) + {\cal O}(\epsilon^2) \, ,
\nonumber \\ 
\chi({\bf r}+a {\bf e}_{\mu},j) &\rightarrow& \chi({\bf r},\tau) + a {\bf e}_{\mu}  \cdot \nabla \chi({\bf r},\tau) + {\cal O}(a^2) \, ,
\nonumber \\  
\displaystyle \sum_j \epsilon &\to& \int d\tau \, \, {\rm and} \displaystyle \sum_{{\bf r}} a^2 \to \int d^2{\bf r} \, ,
\nonumber
\eea 
where $a$ is the lattice spacing.
We have not written the time and space variations of the field
$\chi_2$ because this field is massive and therefore the low energy
physics is dominated by the $\chi^2_2$ term. Because of that we
can integrate out the $\chi_2$ term in perturbation theory in $\lambda$
in order to generate an effective field theory for the field $\chi_1$.
In this case we obtain a relativistic-like field theory of the form:
\bea
{\cal S}_{\rm{eff}} &=& \int d^2\vec{r} d\tau \left\{ \frac{K}{2} \left[(\nabla
  \chi_1)^2 + (\partial_{\tau} \chi_1)^2  \right] \right.
\nonumber
\\
&-& \left. \gamma \cos(4 \pi \chi_1)\right\} \, ,
\label{seffchi1}
\eea
where, $K$ is the stiffness, and $\gamma = \pi \lambda^2/(2 \alpha)$ 
(we have set the velocity of propagation of the field equals to one, for simplicity). Eq.~(\ref{seffchi1})
describes a Sine-Gordon problem in $2+1$ dimensions. Notice that in $2+1$ dimensions 
the stiffness has dimensions of energy and, since the only energy scale in this problem
is the hopping $t$ (since $g \to 0$), one concludes that $K \propto t$. ~\cite{note}
Since we have 
assumed that $\gamma \ll 1$ we can study this problem using a renormalization
group argument \cite{Polyakov,kosterlitz}, that is, we study the relevance of the cosine operator
perturbatively by integrating high energy modes in a shell between 
$\Lambda$ and $\Lambda + d\Lambda$ where $\Lambda$ is the ultraviolet
cut-off of the theory. In doing that the coupling $\gamma$ renormalizes
as \cite{Polyakov,kosterlitz}:
\bea
\frac{d \gamma}{d \ell} = \gamma^2 \, ,
\eea
where $d \ell = d\Lambda/\Lambda$. Hence, $\gamma$ is a marginally relevant
coupling indicating that the field $\chi_1$ is ``frozen'' 
at the minima of the potential in (\ref{seffchi1}), that is, 
at $\langle \chi_1 \rangle = \pm 1/4$ and a gap opens in the spectrum
of $\chi_1$ (thus, both $\chi_1$ and $\chi_2$ are massive). 
The above result could be also derived directly from (\ref{effactionfreezed})
by minimizing the potential energy in order to find:
\beq
\langle \chi_1 \rangle = \mp 1/4 ~;~ \langle \chi_2 \rangle  = 
\pm  { \pi \lambda \over \alpha} \cos(2 \pi \langle \chi_2 \rangle) .
\eeq
Thus, the energetic cost of instanton configurations in this systems 
forces it to freeze on one of the
configurations that minimizes the potential. As mentioned before,
the resulting ground-states present the  N\'eel
order. Thus, we have shown by a mechanism of
order-by-disorder that the classical degeneracy is lifted by quantum
fluctuations and select a state with N\'eel order.
This conclusion is consistent with numerical 
simulations \cite{others,olav}.

To understand what is the effect of a finite temperature, it is better to
come back to the discrete action (\ref{effactionfreezed}). We fist notice that the coarse
grained system has effective $Z_2 \otimes U(1)$ symmetry, corresponding
to changing $\chi_2 \to - \chi_2$ and $\chi_1 \to \chi_1 + n/2$ (where $n$ is
an integer). The ground state occurs with the simultaneous breaking of
the two symmetries leading to a flat configuration of the fields. As the
temperature is increased we expect the restoration of the $Z_2$ symmetry
and a rough phase with algebraic correlations for the $\chi_1$ field. The effective
two-dimensional action at finite temperatures has the form:
\bea
{\cal S} &=& \beta \int d^2\vec{r} d\tau \left\{ \frac{1}{2} \left[K (\nabla \chi_1)^2 +\alpha
  (\chi_2)^2 \right] \right. \nonumber \\
 &+& \left.  \lambda \sin(2 \pi \chi_1) \sin(2 \pi \chi_2) - \gamma \cos(4 \pi \chi_1) \right\}   \, . 
 \eea 
where $\beta=1/T$ is the inverse temperature. Clearly, the phase transition from
a disordered high temperature phase to the ordered low temperature phase
occurs with the breaking of the $Z_2 \otimes U(1)$ symmetry. As usual, the
critical temperature, $T_c$, is proportional to the stiffness $K$ and therefore, 
$T_c \propto K \propto t$, consistent with the statement that $t$ is the
only physical energy scale in this problem.

Equation (\ref{seffchi1}) also allows us to see explicitly how defects that are created 
over such ground state are confined by following Polyakov's 
argument \cite{Polyakov}. 
While this is a result that one could easily
anticipated from the nature of the ground state itself, its
description in terms of a gauge theory will be useful in 
higher dimensional models discussed in the next section. 
As we argued above,
and as also considered in similar systems on Ref.~[\onlinecite{RS}], we look for
an effective action of the field $\chi_1$ which couples to the
matter field (the defects). Integrating over the massive field
$\chi_2$ and expanding $\tilde{\chi} = \chi_1 - <\chi_1>$ in
Eq.\eqref{seffchi1} we obtain the effective action:
\beq
{\cal S} = \int d\vec{r} d\tau \left\{ \frac{K}{2} \left[(\nabla \tilde{\chi})^2 + 
  (\partial_{\tau} \tilde{\chi})^2 \right] + m^2 ~ \tilde{\chi}^2 \right\} \, ,
  \label{massf}
\eeq
where $m^2 \approx 2 \pi^2 \gamma$. 
These fluctuations are then massive and decay over a length scale
$\sim 1/m$ reflecting the fact that two-component Coulomb gas has 
Debye screening. This is also the confinement scale which defines
the string tension $\sigma$ of the defects, that is, the
confining potential has the form: $V(R) = \sigma R$ where $\sigma \sim m$. 
This argument follows from Polyakov's analysis \cite{Polyakov} of the standard
compact QED in $2+1$ dimensions which gives the same effective
action as in Eq.\eqref{massf}. 
At finite temperature, as long as the system remains in the flat phase,
the ionic defects will be confined. Increasing further the temperature
will make the system enter the rough phase, at which the effective theory
can be described by a massless theory for the field $\chi_1$ with $U(1)$ symmetry 
\cite{Lieb}. Creation operators of ionic defects correspond to vertex operators
of the dual fields (or dislocations in the surface roughening language) which
now have algebraic decaying correlation functions.

\section{Three-dimensional models}
\label{threedimensions}

In $3+1$ dimensions the situation is more interesting since there is a
confining-deconfining transition in the normal compact QED at $T=0$. It
is then in the cubic and pyrochlore lattice where we do have a
chance to see deconfined ionic excitations even at $T=0$. This argument has been
used recently by Hermele {\it et al.} in the context of the spin ice
on the pyroclore lattice \cite{HFB}.
The simplest three-dimensional lattice where we can apply this description
is the cubic lattice. On this lattice, we can in fact write
the Hamiltonian with precisely the same form as in $2+1$ dimensions.
The main difference with the standard QED is again the term in
Eq.\eqref{lagrange} which penalizes configurations with $E_{\mu} \neq 0,1$. 
In Ref.~[\onlinecite{HFB}], numerical evidence is provided that due
to this term in Eq.\eqref{lagrange} the defects are deconfined, which is
to say that in the current model corresponds to a quantum disordered
phase of ice (proton liquid). However, this analysis is based on a ``naive''
continuum limit of a lattice model and therefore is only applicable
deep in the deconfined phase. It is known from early studies of 
compact QED in $3+1$ dimensions that this theory has a
deconfinement-confinement
quantum phase transition at a critical value of the coupling constant
$t$,  the resonance amplitude \cite{lautrup,kogut82}. 
Qualitatively this phase transition is driven by the proliferation
of ``monopole loops''. Notice that the theory described here, being
perturbative in $t$, cannot describe the details of the quantum critical point
that occurs at finite values of $t$.
Nevertheless, these general arguments do predict the existence of a deconfinement-confinement quantum transition at finite $t=t_c$ and $T=0$ \cite{fradkin}. The actual location of this quantum critical point may be controlled by additional operators not included here, such as the short range part of the dipole interaction, which affect the ``electric'' terms but leave the ``magnetic'' (flip) terms untouched. It is likely that the model we are considering here, which has a single coupling constant $t$, with the electric terms playing the role of a constraint, may be in the deconfined (or ``Maxwell'') phase.

The proton phases of ice are characterized by the energy required
to separate two defects with opposite sign. In the confined phase (the
ordered phase of ice) 
the energy grows linearly with the separation between the ions.
Therefore, in this phase it is not possible to separate the H$_3$O$^+$ and
HO$^-$ ions at arbitrary distance 
with the application of an external electric field. This state is a 
dielectric with a finite polarizability and it is an insulator.
However, in the deconfined phase the effective interaction between the
ions obeys $V(R) \propto 1/R$ and therefore decreases with the
distance between the ions. Thus, in this case an applied electric field
can separate the two H$^+_3$O and HO$^-$ ions at arbitrary distance 
leading to a conducting state. Hence, the phase transition is 
a metal-insulator transition \cite{mott} and therefore the conductivity
of ice should behave quite different in the two different sides of the
transition. We should note that in the metallic phase the conduction
is due to the collective motion of protons. This is a correlated metal. 

\begin{figure}[htb]
\psfrag{T}{$T$}
\psfrag{Tc}{$T_c$}
\psfrag{t}{$t$}
\psfrag{tc}{$t_c$}
\psfrag{QCP}{QCP}
\psfrag{Deconfined}{Deconfined Disordered}
\psfrag{Plasma}{Proton Plasma}
\psfrag{Confined}{Confined Ordered}
\psfrag{Insulator}{Proton Insulator}
\begin{center}
\subfigure[]{\includegraphics[width=0.35\textwidth,keepaspectratio]{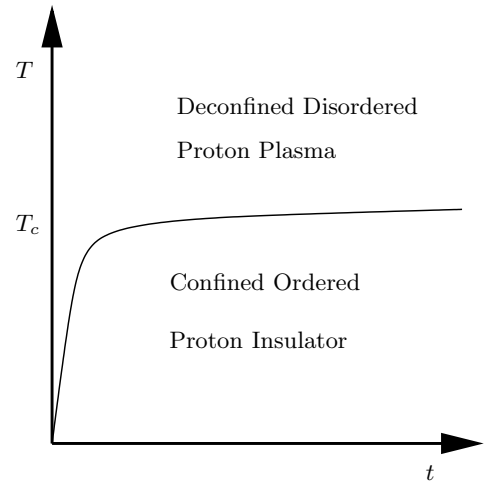}}
\\
\subfigure[]{\includegraphics[width=0.35\textwidth,keepaspectratio]{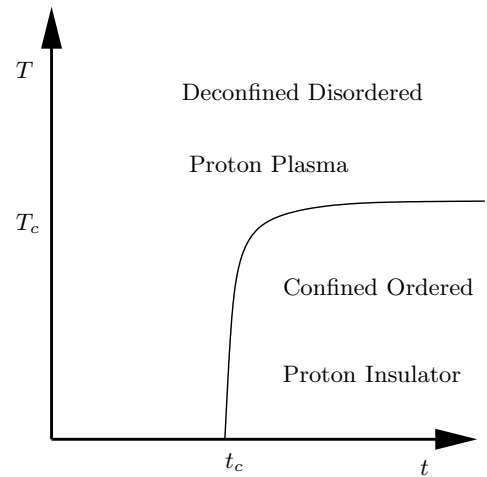}}
\end{center}
\caption{Phase diagram for protons in neutral ice. (a) Planar ice; (b) Three
  dimensional ice.}
\label{phase_diagram}
\end{figure}

In Fig.\ref{phase_diagram} we depict the phase diagram of ice as
a function of temperature and the proton hopping energy $t$. 
In Fig.\ref{phase_diagram}(a) we show the phase diagram for 
planar ice ($2+1$ dimensions) for which we have shown that there is no
deconfining transition as a function of $t$ at $T=0$. 
As we argued in the last section, there will be, however,
a phase transition at finite temperature $T_c$ ($T_c \propto t$ for $t \to
0$)  between the disordered metallic phase and the
ordered insulating phase.
In the case at hand, the arguments presented above indicate the presence of a
line of second order phase transition \cite{conjecture}. As we are going to
see in the next Section, the coupling to phonons can change
the transition to first order. 
In the case of the pyroclore or cubic lattice ($3+1$ dimensions) 
we have argued that there is
a confining deconfining transition as a function of $t$, 
as shown in Fig.\ref{phase_diagram}(b). In this case, there is
a quantum critical point (QCP) at some value of $t$ (say, $t_c$).

\section{Coupling to lattice vibrations}
\label{cphonons}

So far we have not discussed the problem of lattice
vibrations in ice. At finite temperatures one expects that the thermal
motion of the O atoms to affect the properties of the H atoms.
Vibrations must be important since the mass of the H atoms are
just 16 times smaller than the O atoms and therefore the
Born-Oppenheimer approximation is not guaranteed. In order to
incorporate the recoil of the O atoms due to the H motion 
we assume that the phonon coordinates
are coupled to the local proton density. In this case, 
one has to add an extra term to the Hamiltonian of the form: 
\bea H_P = &-&
\kappa \sum_{i,\alpha,\sigma} n_{i,\alpha,\sigma} q_{i,\alpha,a} 
\nn
\\ &+& \sum_{i,\alpha,a} \left(\frac{P_{i,\alpha,a}^2}{2 M} + \frac{M
\omega_{i,a}
 q_{i,\alpha,a}^2}{2} \, ,
\right) \label{phonons} 
\eea 
where $q_{i,\alpha,a}$ are the local
phonon coordinates, $P_{i,\alpha,a}$ their canonical momentum, $\omega_{i,a}$ 
are the phonon frequencies, and $M$ the ion mass. Eq.\eqref{phonons} can be
reduced to the two level system in the same way Eq.\eqref{hamileff} is
reduced to Eq.\eqref{ising}. Using Eq.\eqref{transformation} and
defining the relative and center of mass coordinates: 
\begin{eqnarray}
x_{i,a} &=& q_{i,1,a}-q_{i,2,a} \nonumber \, ,
\\
X_{i,a} &=& \frac{q_{i,1,a}+q_{i,2,a}}{2} \, ,
\end{eqnarray}
the Hamiltonian Eq.\eqref{phonons} reduces to 
\bea 
H_P &=& - \kappa \sum_{i,a} S_{i}^z x_{i,a} + \sum_{i,a}
\left(\frac{p_{i,a}^2}{M} + \frac{M \omega_{i,a} x_{i,a}^2}{8} \right.
\nonumber
\\
&+& \left. \frac{\Pi_{i,a}^2}{4 M} + \frac{M \omega_{i,a} X_{i,a}^2}{4} \right)
\, ,
\label{proton_phonon}
\eea 
where $p_i$ is conjugated to $x_i$ and $\Pi_i$ is conjugated
to $X_i$. Notice that the center of mass coordinate decouples
from the proton motion in this case. 

It is interesting to rewrite the full Hamiltonian of the problem
in terms of the pseudo-spin operators and standard creation $b^{\dag}_i$  
and annihilation $b_i$ 
operator for the phonons ($[b_i,b^{\dag}_j]=\delta_{ij}$). 
Using Eq.\eqref{ising} and proton-phonon coupling term Eq.\eqref{proton_phonon} 
we find:
\begin{eqnarray}
H &=& J \sum_{\langle i,j \rangle} S^z_i S^z_j - \Gamma \sum_i S^x_i
+ \sum_{i,a} \lambda_{i,a} S^z_i (b_{i,a} + b_{i,a}^{\dag}) 
\nonumber
\\
&+&  \sum_{i,a} \omega_{i,a} b^{\dag}_{i,a} b_{i,a} 
\label{ising_phonon}
\end{eqnarray}
where $\lambda_{i,a} = \kappa/\sqrt{M \omega_{i,a}}$. Eq.\eqref{ising_phonon}
describes an Ising model in a transverse field coupled to a dissipative environment
{\it \`a la} Caldeira-Leggett \cite{caldeira}. This model has been
studied recently in the context of quantum phase transitions in metallic
magnetic systems \cite{griffiths,leticia} and has very interesting
properties.  

Let us consider the classical case of $\Gamma \ll J$ when the
ice rules are obeyed. In this case the problem can be solved in
the basis of $S^z$: $S^z_j |\sigma_j\rangle = \sigma_j |\sigma_j\rangle$.
It is easy to see that the problem can be diagonalized
by shifting operators:
\bea
B_{j,a} &=& b_{j,a} + \frac{\lambda_{j,a} \sigma_j}{\omega_{j,a}}\, ,
\nonumber
\\
B^{\dag}_{j,a} &=&  b^{\dag}_{j,a} + \frac{\lambda_{j,a}
  \sigma_j}{\omega_{j,a}} \, ,
\eea
and the energy of the problem is given by:
\bea
E[\{\sigma,n\}] &=& J \sum_{\langle i,j \rangle} \sigma_i \sigma_j
\nonumber
\\
&+& \sum_{i,a} \left[\omega_{i,a} \left(n_{i,a} + \frac{1}{2}\right) -
\frac{\lambda_{i,a}^2}{\omega_{i,a}}\right]
\eea
where the first term is just the energy of the classical state and 
$n_{i,a}$ is the phonon number. 
So, the phonon frequencies remain the
same but the atoms positions are shifted by $\delta x_{i,a}
\propto \kappa \sigma_i/(M \omega_{i,a}^2)$. Since there is no
proton order when $\Gamma=0$ the average lattice shift is zero.
However, for a small value of $\Gamma$ we see from Fig.~\ref{phase_diagram}
that $\sigma_i$ acquires an average expectation value leading to
an overall shift in the atom positions. At finite temperature 
this shift will lead to a discontinuity in the specific heat as
it is well known in the case of cooperative transitions of this
sort \cite{cooptrans}. Thus, the phase diagram 
second order phase transition of Fig.\ref{phase_diagram} can
be modified substantially. In $2+1$ dimensions the finite temperature
phase transition can change completely to first order while in $3+1$
dimensions a tricritical point must appear at some temperature $T^*$
so that for $T<T^*$ the phase transition is second order and for
$T>T^*$ the transition becomes first order.

\section{Conclusions and Experimental Realizations}
\label{conclusions}

In this paper we have considered the phase diagram of
protons in ice. We have given a detailed picture of 
two-dimensional ice and provide a qualitative picture of the 
three-dimensional problem such as
the pyrochlore lattice. 
We have studied the order-by-disorder effect and
ground-state selection by quantum fluctuations, which in this
case is the  tunneling of protons 
between the two sites in a Hydrogen bond.

In the absence of quantum tunneling, the number of low energy
states grows exponentially with the size of the system and the
entropy of the system is macroscopic. In this classical background,
excited states corresponding to ionic defects can be created and
separated without further cost in energy. As we showed, 
the presence of quantum fluctuations change considerably
this scenario. The degeneracy of the system is lifted by the
order-by-disorder mechanism leading to a well defined ground state.
The technique used to understand the role of
quantum fluctuations is based on a mapping of the lattice problem 
onto a gauge theory. 

Within the gauge theory one can describe
the behavior of ionic defects as confined or deconfined depending
whether the ground state of the system is ordered or disordered, 
respectively.  In a deconfined state the ionic defects behave 
like as in a correlated metal, while in the confined phase the system 
behaves as in an insulator. For the two-dimensional model, 
quantum fluctuations select a a
two-fold degenerate antiferroelectric ground-state over which
ionic excitations are confined. The situation is different in the
three-dimensional case where topological excitations can lead
to a deconfined disordered state even at $T=0$. In such
case, the interplay of quantum fluctuations and the tuning of the
microscopic tunneling parameter is expected to give rise to a
confinement-deconfinement transition for ionic defects, with their
corresponding dramatic effects in the dielectric properties of the
material. 

Furthermore, we have argued that 
at finite temperatures there will be a phase
transition between ordered (confined) and disordered (deconfined) ice.
The nature of this phase transition depends on the fugacity associated
with the topological defects (monopoles) and can be either first or
second order. In the pyroclore lattice the most likely scenario is that
there is a tricritical point separating a first order from a 
second order phase transition (see Fig.\ref{phase_diagram}(b)) that
should be observable. While similar problems have
already being addressed in terms of spin models \cite{MS, HFB}, the ice
scenario brings a new perspective to this kind of phenomenology
due the more accessible experimental possibilities than in
frustrated spins systems \cite{fulde}.

It is clear from Fig.\ref{phase_diagram}(b) that the way to tune
the phase transition is by changing the value of the proton hopping
energy $t$. The value of $t$ depends on the overlap of the 
proton wavefunction between the two sites in the Hydrogen bond.
Since the mass of the proton is large its wavefunction is more
localized than in the electronic case and we expect 
that $t$ is exponentially sensitive on the changes in the distance
between O atoms. One clear way to experimentally change $t$ is then
by applying hydrostatic pressure $P$ to an ice crystal at fixed
temperature $T$, starting from the disordered phase, and measuring
the ice conductivity or dielectric response as a function of pressure.
One expects that the proton ordering discussed here occurs
at lower pressures than the well-known structural phase transitions
of ice \cite{hobbs}.

Another way to change the O-O distance
is by using ``chemical pressure'' \cite{kawada}. One can dope water with salts 
like KOH that while in solution become K$^+$ and HO$^-$. 
When in the solid phase the K$^+$ ion becomes trapped into the ``cages'' of
O atoms while the HO$^-$ ions are assimilated into the ice lattice.  
Thus, in this case, doping also introduces charges into the lattice
structure and the ground state is no longer neutral. Furthermore,
since the K$^+$ ions spread randomly over the ice lattice, this kind
of doping also introduces disorder in the system.
The attractive interaction between the positive K$^+$ ions and
the negative O$^{-2}$ atoms of the ice structure lead to a local decrease
of the volume of ice ``cages'', to an average decrease in the
H-bond distance, and to an average increase in $t$. So, in first
approximation the introduction of KOH is somewhat equivalent to
the horizontal axis in Fig.\ref{phase_diagram}. 
However, because there is introduction of charges and disorder 
in the system, the horizontal axis in Fig.\ref{phase_diagram}
is only roughly the concentration of KOH.

Doping experiments with KOH 
have been performed more than $20$ years ago \cite{kawada} 
in order to understand
the famous $72$ K anomaly observed in the temperature 
dependence of the specific heat of undoped ice \cite{dopspec}. 
The specific heat of pure ice at low temperatures increases with
temperature as $T^3$ which is the characteristic of phonons in
the material. However, it has been known for a long time that
hexagonal ice (ice-Ih) has a bump in the specific heat at $T \approx 72$ K that 
was a theoretical mystery \cite{onsager}. By doping pure ice with KOH it was shown
by Kawada in the 1970's that the specific heat bump is actually a
very slow phase transition into an ordered proton state that 
was called ice XI \cite{kawada,structure}. Permittivity experiments
in doped ice have confirmed this scenario \cite{oguro}. 
Specific heat measurements in KOH doped ice showed a strong and 
highly hysteretic
first order phase transition with a substantial loss of entropy
in the low temperature phase \cite{dopspec}. Neutron scattering
experiments on single crystals of ice have confirmed the ordering
transition \cite{jackson,D2O}. While the critical
temperature is weakly dependent on the amount of KOH, the loss
of entropy is dependent on the KOH concentration. Furthermore,
in accordance with our discussion in the previous section, the
first order phase transition is associated with lattice 
distortions as seen in recent neutron scattering experiments
\cite{phonons}. 

Hence, our theory provides a possible theoretical
explanation for the phase transition between ice-Ih and
ice-XI. 
We believe that the first order phase transition observed
in the experiments is the one
described in Fig.\ref{phase_diagram}(b) and it would be
very interesting to find out whether 
the quantum critical point can be studied by further doping of KOH
or by application of pressure. 
Our theory indicates that at $T \to 0$ and $t<t_c$ the protons in ice-Ih 
would make an  unique state of matter, namely, a quantum proton liquid,
with deconfined ion excitations. 

\acknowledgments
This work was started at least 8 years ago and we had the opportunity
for illuminating conversation with many colleagues over the years.
We would like to thank particularly  A.~Auerbach, W.~Beyermann, 
A.~Bishop, D.~K.~Campbell, S.~Chakravarty, M.~P.~A. Fisher, P.~Fulde, M.~Gelfand, L.~Ioffe,  
D.~MacLaughlin, R. Moessner, C.~Mudry, C.~Nayak, S.~Sachdev, S.~L. Sondhi, and H.~Tom. 
P. Pujol would like to thank the quantum condensed matter visitor's program at the Physics Department of Boston University, were part of this work was carried out.  This work was
partially supported by the National Science Foundation through grants DMR-0343790 at Boston University (AHCN) and DMR-0442537 at the University of Illinois (EF).


\begin{thebibliography}{99}
\bibitem{note} The actual value of the stiffness $K$ cannot be directly calculated from
the naive coarse grained theory in the continuum limit because its value is renormalized
by fluctuations.
\bibitem{conjecture}In Ref.~[\onlinecite{olav}] it is conjectured that this transition is of
first order, instead of second order.
\end{thebibliography}
\end{document}